# ESTIMACIÓN DE LA TRAYECTORIA DE CORONAVIRUS COVID-19 ADHERIDAS A GOTITAS RESPIRATORIAS PROYECTADOS HORIZONTALMENTE, CONSIDERANDO LA ALTITUD GEOGRÁFICA

Julio Warthon[1]*, Amanda Olarte[1], Bruce Warthon[1]
[1] Laboratorio de Físico Química de la Atmósfera, Universidad Nacional de San Antonio Abad del Cusco, 08000 Cusco, Perú.
Correspondencia: julio.warthon@unsaac.edu.pe

**RESUMEN**

El presente estudio trata sobre la estimación de la trayectoria de coronavirus COVID-19 adheridas a gotitas respiratorias proyectados horizontalmente, considerando la altitud geográfica. El *tamaño* de los virus y de las gotitas respiratorias constituye el factor que determina la trayectoria de las micropartículas en un medio viscoso como el aire; para este propósito se ha realizado una comparación gráfica de los diámetros y masas de las micropartículas que se producen en la actividad respiratoria. La estimación del movimiento vertical de las micropartículas a través del aire se sustenta en la Ley de Stokes, se determinó que las gotitas respiratorias menores a 10 µm de diámetro tienen velocidades terminales muy pequeñas, en la práctica *se encuentran flotando* durante breves segundos antes de evaporarse en el aire; respecto al desplazamiento horizontal de las gotitas respiratorias se ha utilizado fotogramas de Scharfman et al. para determinar su alcance. En el caso de un estornudo, las gotitas respiratorias pueden llegar a 1.65 m de distancia en 1 s frenándose rápidamente hasta llegar a 1.71 m en 2 segundos, luego se hizo un análisis del efecto de la altitud geográfica en el movimiento de las microgotas, determinando un cambio mínimo en las variables cinemáticas.

**PALABRAS CLAVE:** Estimación, trayectoria, velocidad terminal, alcance, gotita respiratoria, coronavirus, COVID-19, Stokes, tamaño, diámetro, masa, altitud.

## 1. Introducción

La Física nos permite realizar una aproximación importante de la dinámica de micropartículas, como de los coronavirus COVID-19, cuyo diámetro promedio es 0.1 µm ó 100 nm (Barón, Y.M. et al, 2020); y gotitas que resultan del habla de diámetros entre 3.5 a 5 µm (Morawska L. et al, 2009), un rango intermedio son gotitas respiratorias con diámetros entre 5 a 10 µm (World Health Organization, 2020), y gotas de Flügge con diámetros iguales o mayores a 100 µm. En la exhalación se emite $CO_2$, O, N y vapor de agua, en el caso de estornudos o tos se emite micropartículas de agua de diferentes diámetros. La transmisión por COVID-19 se debe en una parte a que estos virus se adhieren a gotitas respiratorias de 5 – 10 µm que resultan de la actividad respiratoria (World Health Organization, 2020), asimismo hay gotas de Flügge con diámetros igual ó mayor a 100 µm, con carga viral mayor. La diferencia de diámetros entre un coronavirus y una gotita respiratoria es de alrededor de 50 a 100 veces, es decir el virus es mucho menor en tamaño y masa que una gotita respiratoria, por consiguiente, el virus no se puede considerar "pesado" respecto al medio de transporte que son las gotitas respiratorias; en el caso de realizar una comparación entre el nuevo coronavirus respecto a otros virus de la misma familia se puede observar que sus masas no difieren sustancialmente, esta diferencia de masas y diámetros no influye en su desplazamiento respecto a distancias alrededor de 1 m.

Como parte de la metodología se ha realizado inicialmente una comparación gráfica de las diferentes micropartículas que se producen en la exhalación y en el estornudo o tos de una persona con síntomas respiratorios, luego se ha empleado la Ley de Stokes, asumiendo que el aire es un fluido viscoso de régimen laminar de bajo número de Reynolds, este caso se aproxima a un ambiente aislado en el cual no hay ventilación ni movimiento de personas; en el caso de que se considere micro flujos de aire debido a las variaciones de presiones, temperaturas y del movimiento de personas; el análisis es complejo debido a variables estocásticas que se enmarcan en la dinámica no lineal. El aire en principio es un medio homogéneo donde se producen procesos físico-químicos, en este medio viscoso se dan mecanismos de transporte de masa y calor, como caso particular el transporte o desplazamiento de gotitas respiratorias de un punto a otro, los otros componentes de la actividad respiratoria son los aerosoles que son vehículos de transmisión aérea de virus debido a su diámetro que es 35 a 50 veces mayor respecto a un coronavirus COVID-19. Una gotita respiratoria de 10 micras que cae en el aire, experimenta la viscosidad del medio de tal forma que su velocidad terminal sería aproximadamente 3 mms$^{-1}$, es decir, se encuentra flotando desde la perspectiva de un observador en reposo. Según la OMS estas gotitas respiratorias pueden transferirse entre personas cuando están en contacto cercano (dentro de 1 m) con otras personas con síntomas respiratorios (World Health Organization, 2020). Se ha indicado que el virus es "pesado", sin embargo, los resultados obtenidos en la estimación realizada muestran lo contrario, su tamaño y masa no influyen en la cinemática de las gotitas respiratorias a las cuales se adhieren, no describen parábolas en el aire debido a que la viscosidad restringe su movimiento (Ley de Stokes). Los fotogramas experimentales de la investigación realizada por Scharfman (Scharfman, BE, Techet, AH, Bush, JWM et al, 2016) han permitido realizar la estimación del movimiento de gotitas respiratorias.

## 2. Metodología

El procedimiento ha consistido en visualizar gráficamente los tamaños y diámetros de las diferentes partículas que se producen en la exhalación, estornudo ó tos de una persona con síntomas respiratorios; se ha realizado un cálculo de las masas de las micropartículas en estudio; este procedimiento ha permitido diferenciar tamaños y masas de los coronavirus de 0.1 µm de diámetro y los medios de transporte: gotitas resultantes del habla de 3.5 a 5 µm de las gotitas respiratorias de 5 a 10 µm y de una gota de de Flügge de 100 µm de diámetro. Como se demuestra en el presente estudio, no es igual el movimiento de un virus de 0.1µm a una gota de 100 µm en un medio viscoso.

El objetivo ha sido estimar la trayectoria de coronavirus COVID-19 adheridas a gotitas respiratorias y proyectados horizontalmente, considerando la altitud geográfica, para lograr este propósito se ha recurrido a las leyes de la dinámica Newtoniana (Tipler P.A., Mosca G. 2010), la Ley de Stokes (Sears F.W., Zemansky M.W., Young H.D., Freedman R.A. 1999) y de fotogramas experimentales de Scharfman. Se ha realizado el análisis del movimiento en dos dimensiones **xy**, en cada eje se han obtenido las ecuaciones del movimiento y sus soluciones respectivas, posteriormente se ha determinado las constantes respectivas. Las ecuaciones describen el movimiento de las gotitas respiratorias considerando la inamovilidad del aire y su viscosidad, asumiendo régimen laminar de bajo número de Reynolds. Los gráficos de las ecuaciones cinemáticas para los casos estudiados describen la estimación del movimiento.

Para estimar la dinámica de las micropartículas en un punto geográfico distinto a nivel del mar, se requiere el valor de la viscosidad en la altitud el cual varía debido a que la cantidad de masa por unidad de volumen es menor respecto al nivel del mar, para este propósito se ha utilizado una calculadora denominada *1976 Standard Atmosphere Calculator - Digital Dutch*, en este programa informático se ingresa únicamente la altitud del lugar y la temperatura, como referencia se ha elegido una altitud de 3000 m.s.n.m. y una temperatura de 20 ºC; las ciudades o poblados sobre la Tierra se encuentran entre los 0 y los 5000 m.s.n.m., de aquí la altitud elegida es representativa.

**Estimación del movimiento vertical**

Una partícula esférica que desciende verticalmente está expuesta a tres fuerzas: la fuerza viscosa del aire ($F_v$), la fuerza de empuje ($E$) y el peso de la partícula ($w$). La Ley de Stokes permite medir la fuerza de fricción que experimenta una partícula esférica pequeña en un fluido viscoso y en régimen laminar (Sears F.W., Zemansky M.W., Young H.D., Freedman R.A, 1999).

$$F_v = 3\pi \eta d v_t$$

$\eta$: viscosidad del fluido
r: radio de la esfera
$d$: diámetro de la esfera
$v_t$: velocidad terminal de la esfera
$\rho$: densidad del fluido
$\rho^{\cdot}$: densidad de la esfera

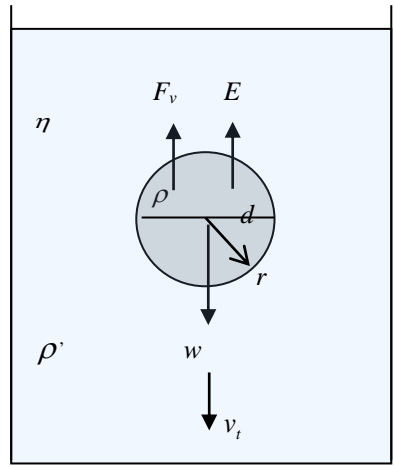

Figura 1 Fuerzas que actúan sobre una partícula en un fluido viscoso

La ecuación del movimiento en el eje **y** (dirección vertical) se da mediante la segunda Ley de Newton, considerando el empuje y la fuerza viscosa en la dirección vertical positiva y el peso de la partícula (gotita respiratoria, coronavirus COVID-19) en la dirección opuesta al anterior:

$$m\frac{d^2 y}{dt^2} = \frac{1}{6}\pi . d^3 \rho^{\cdot} g + 3\pi \eta d v_y + (-mg)$$

La velocidad terminal de gotitas respiratorias y de coronavirus COVID-19 en el aire se da cuando las micropartículas consiguen una velocidad de descenso invariable, luego la aceleración en el eje **y** es cero:

$$v_t = \frac{1}{18} . \frac{d^2 g}{\eta}(\rho - \rho^{\cdot})$$

El desplazamiento vertical (hacia abajo) es:

$$y = \frac{1}{18} . \frac{d^2 g}{\eta}(\rho - \rho^{\cdot}) . t$$

**3 Resultados y discusión**

**3.1 Comparación gráfica de los diámetros de las micropartículas expulsadas en la exhalación o estornudo de una persona con síntomas respiratorios**

El movimiento de los cuerpos es un fenómeno físico en principio, que se puede medir a partir de leyes generales del movimiento, sin embargo el factor tamaño influye en su movimiento. La mecánica clásica permite evaluar el movimiento de los cuerpos o partículas en ciertos rangos de longitudes o tamaños y no puede aplicarse a nivel cuántico o atómico de igual forma, incluso en el rango -alrededor del micrómetro- que se realiza el análisis del movimiento de las gotitas de la actividad respiratoria, se deben considerar las magnitudes geométricas; el tamaño nano y micrométrico influyen en el movimiento de las partículas en un medio viscoso como el aire, por esta razón se ha procedido a comparar gráficamente el diámetro de las diferentes partículas que resultan de la actividad respiratoria. En el gráfico realizado se considera el diámetro de las partículas y las diferencias en cada caso; referente a la masa, se ha elaborado una tabla considerando sus diámetros y densidades para cada partícula.

En el siguiente gráfico a escala se muestra la diferencia de los diámetros entre el coronavirus COVID-19 (Barón, Y.M. et al, 2020), las gotitas resultantes del habla (Morawska L. et al, 2009), las gotitas respiratorias (World Health Organization, 2020) y la gota de Flügge.

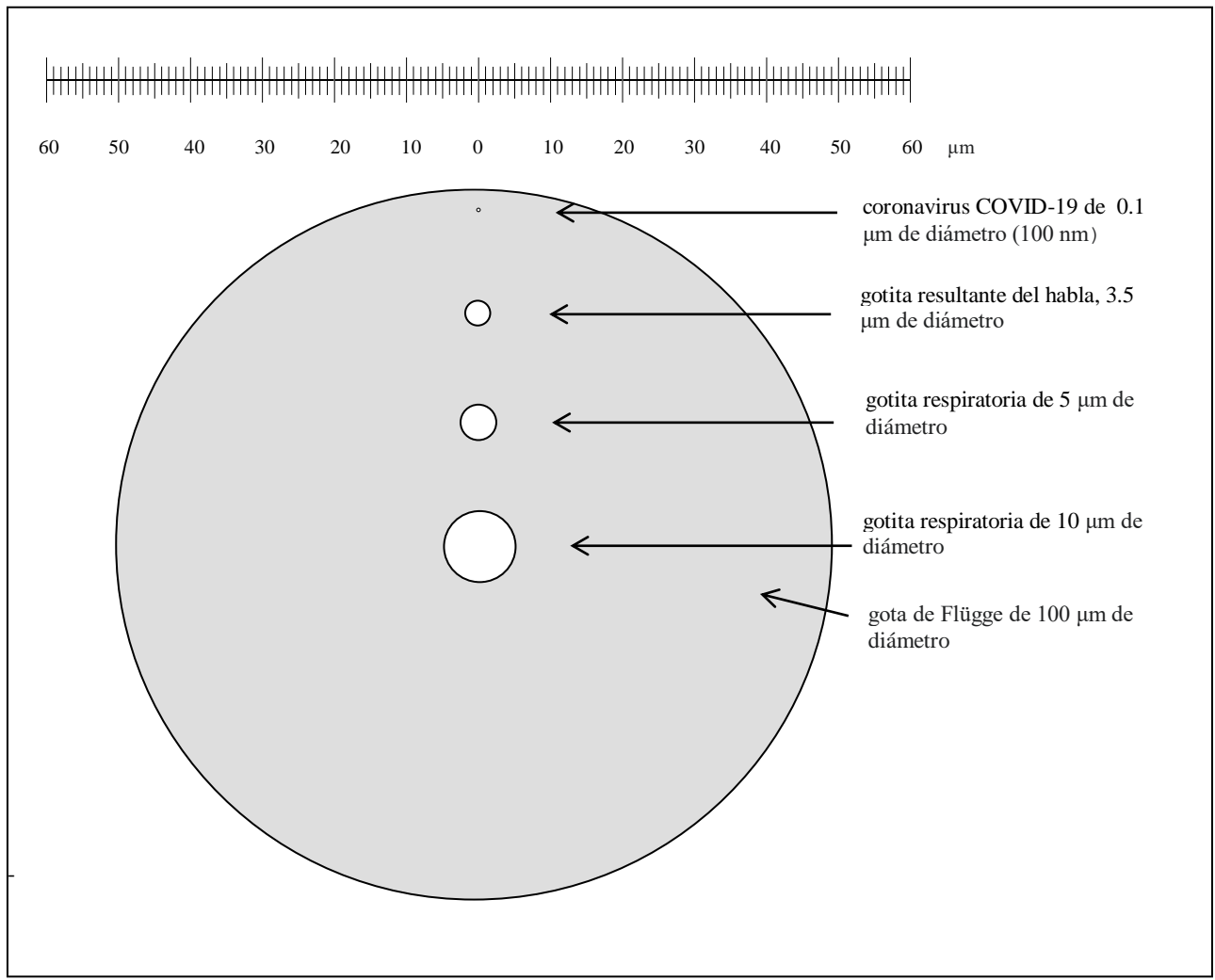

Figura 2 Comparación gráfica de los diámetros de un coronavirus COVID-19, gotita resultante del habla, gotitas respiratorias y una gota de Flügge

Al medir con un micrómetro el grosor de una hoja de papel estándar veremos que es aproximadamente igual a 0.1 mm o 100 μm, es decir la gota más grande que se observa en la figura 2 es igual al grosor de una hoja de papel estándar, nos da una idea de los rangos de magnitud de las microgotas; una gota de Flügge puede ser percibido por un ser humano dependiendo de la edad y capacidad visual, por tanto, las gotitas respiratorias de 5 ó 10 μm no se podrían visualizar de manera individual sino de manera conjunta.

A continuación se muestra una tabla que permite distinguir las masas de diferentes partículas: coronavirus COVID-19, gotita resultante del habla, gotitas respiratorias y gota de Flügge:

Tabla 1 Masas de partículas producto de la actividad respiratoria de un ser humano.

| Partícula | Diámetro medio de la partícula ($\mu m$) | Densidad ($gk/m^3$) | Masa $m(kg)$ |
|---|---|---|---|
| Coronavirus COVID-19 (Barón, Y.M. et al, 2020) | 0.1 | $\approx 10^3$ | $5.24 \times 10^{-19}$ |
| Gotita resultante del habla (Morawska L. et al, 2009) | 3.5 – 5 | $10^3$ | $22.45 - 65.45 \times 10^{-15}$ |
| Gotita respiratoria (World Health Organization, 2020) | 5 | $10^3$ | $65.45 \times 10^{-15}$ |
| Gotita respiratoria (World Health Organization, 2020) | 10 | $10^3$ | $5.23 \times 10^{-13}$ |
| Gota de Flügge | 100 | $10^3$ | $5.23 \times 10^{-10}$ |

La proporción de diámetros es diferente a la proporción de masas, en el caso de un coronavirus COVID-19 y una gotita respiratoria, la relación de diámetros es:

$$\frac{d_{gotita}}{d_{virus}} = \frac{5\mu m}{0.1\mu m} = 50$$

Comparando sus masas:
$$\frac{m_{gotita}}{m_{virus}} = \frac{\rho_{agua}.V_{gotita}}{\rho_{virus}V_{virus}} = \frac{d^3_{gotita}}{d^3_{virus}} = \frac{5^3}{0.1^3} = 125000$$

La masa de una gotita respiratoria de 5 μm de diámetro es aproximadamente 1.25x10$^5$ veces mayor (125000 veces) la masa de un coronavirus COVID-19 de 0.1 μm de diámetro, esta comparación permite afirmar que un virus o varios que tengan este diámetro que se adhieran a la superficie de una gotita respiratoria de 5 ó 10 μm de diámetro, no influirían en la masa total (gotita respiratoria y virus). De aquí podemos concluir que el coronavirus COVID-19 "no es pesado" respecto a las gotitas resultante del habla (rango de aerosol) ni de las gotitas respiratorias o de las gotas de Flügge. Podría considerarse ligeramente de mayor peso respecto a otros virus o similares, pero no influiría en la velocidad terminal de la gotita de transporte; debido a las condiciones termodinámicas del aire una gotita de agua de 10 μm puede conseguir evaporarse en una distancia de 2.1 cm en caída (Rogers R R, p.81,1977), esta distancia recorrida en la caída corresponde a 7 segundos en promedio, concuerda con nuestra observación cotidiana, cuando una película de agua se encuentra en una superficie cualesquiera, se evapora en breves segundos.

### 3.2 Estimación del alcance, velocidad y aceleración de una gotita respiratoria y del nuevo coronavirus.

Se ha asumido el movimiento de una gotita respiratoria proyectada horizontalmente, considerando la resistencia del aire y sin microflujos aleatorios. La fuerza resultante que experimenta la microgota son cuatro, asumiendo que el aire esta inmóvil y que en la dirección vertical logran la velocidad terminal:

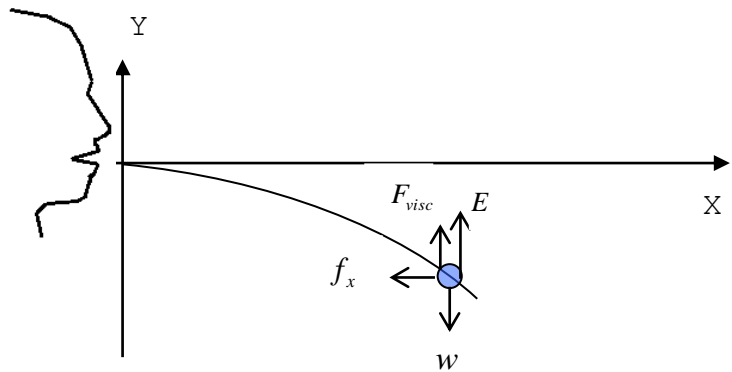

Figura 3 En esta figura se observa las fuerzas que actúan en la dinámica de una microgota de agua

Las ecuaciones respectivas en el plano **xy**, son:
$$m\frac{d^2x}{dt^2} = -f_x \quad (1)$$
$$m\frac{d^2y}{dt^2} = E + F_{visc} - w \quad (2)$$

En el eje **x**, la ecuación diferencial del movimiento es:
$$\frac{d^2x}{dt^2} = -\frac{f_x}{m} \quad (3)$$

La fuerza de fricción $f_x$ es proporcional a la velocidad del objeto en la dirección **x**: $f_x = kv_x$ (Tipler P.A., Mosca G., 2010), donde $k$ es una constante, luego:
$$\frac{d^2x}{dt^2} = -\frac{k}{m}v_x \quad (4)$$

Con el propósito del análisis se tiene una nueva constante: $\alpha = k/m$, luego:
$$\frac{d^2x}{dt^2} = -\alpha v_x \quad (5)$$

De aquí se tienen las tres variables cinemáticas:
$$x = \frac{v_0}{\alpha}\left[1 - e^{-\alpha.t}\right] \quad (6)$$
$$v_x = v_0 e^{-\alpha.t} \quad (7)$$
$$a_x = -\alpha v_0 \, e^{-\alpha t} \quad (8)$$

Donde $v_o$ es la velocidad inicial en la dirección **x**, $\alpha$ es la constante de atenuación del medio relacionada con la viscosidad y la masa de la micropartícula en movimiento; conociendo la velocidad inicial se puede determinar la constante α. A partir de los fotogramas de Scharfman (Scharfman, BE et al, 2016), se ha estimado la velocidad inicial considerando el fotograma donde se observa una distancia de 0.03 m y el tiempo de 0.005 s, de aquí la velocidad inicial es $v_o$=6 $ms^{-1}$; sustituyendo estos valores en la ecuación (6) y mediante un proceso iterativo se obtiene el valor de α=3.5. Las ecuaciones ahora son:

$$x = \frac{6.0}{3.5}\left[1 - e^{-3.5 \cdot t}\right] \quad (9)$$

$$v_x = 6.0\, e^{-3.5 \cdot t} \quad (10)$$

$$a_x = -21 e^{-3.5 \cdot t} \quad (11)$$

Como se puede observar, todas las ecuaciones consideran un factor exponencial que nos indica que las variables cinemáticas de la gotita respiratoria disminuyen en el transcurso del tiempo. Estas ecuaciones pueden describir las trayectorias de las gotitas expulsadas en la actividad respiratoria del ser humano. Los gráficos que se muestran corresponden a gotitas respiratorias de 5 μm de diámetro.

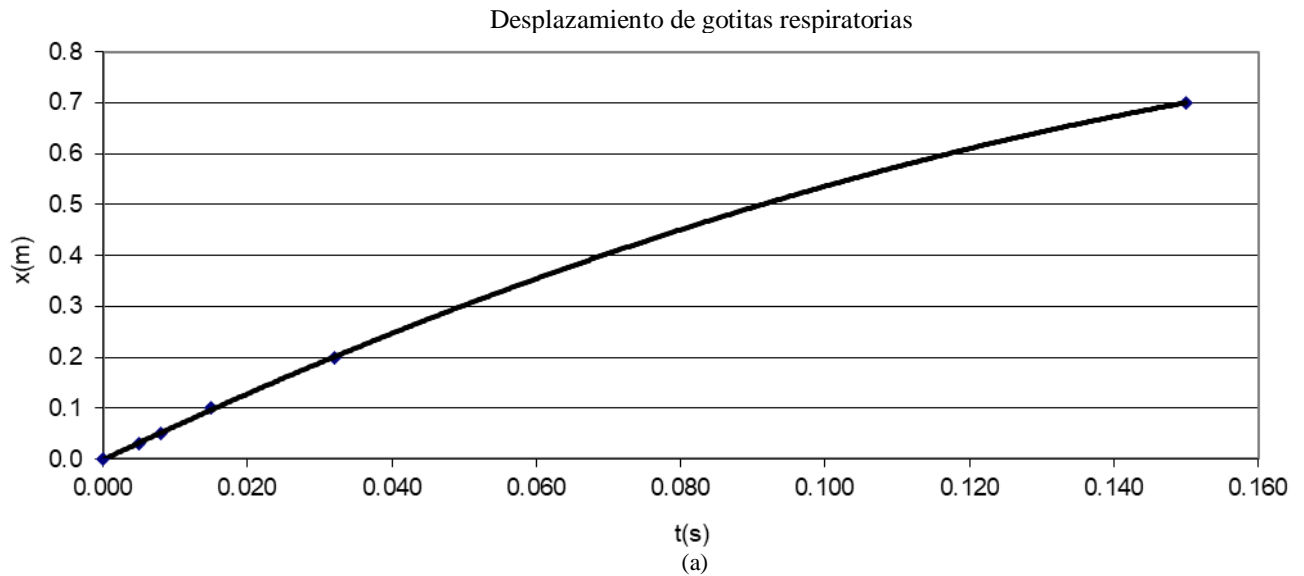

(a)

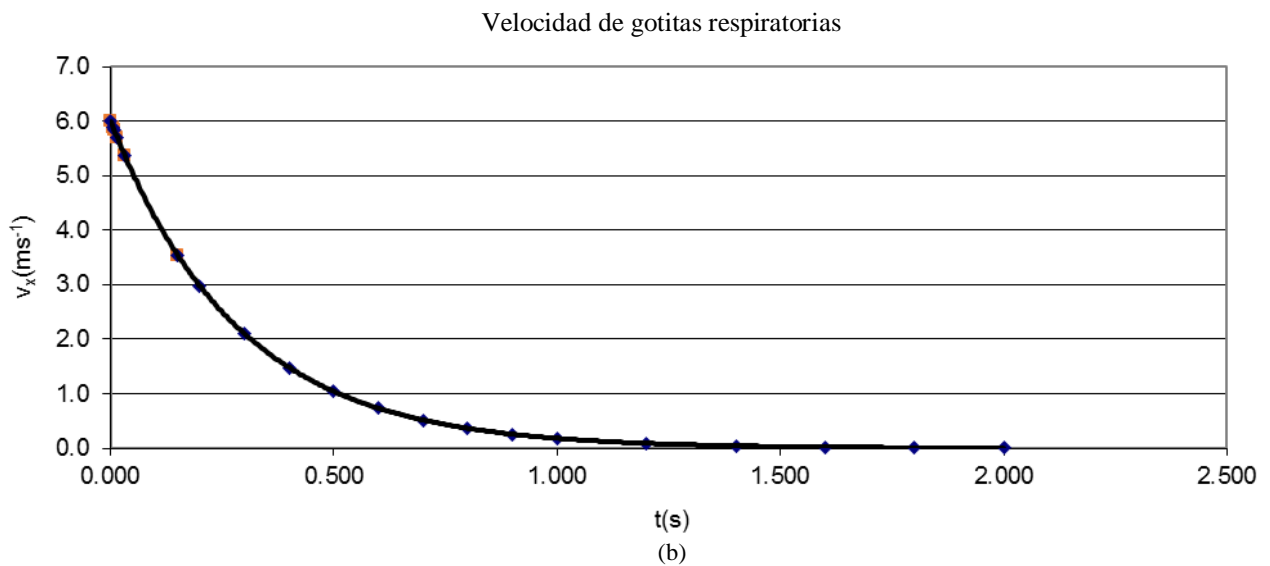

(b)

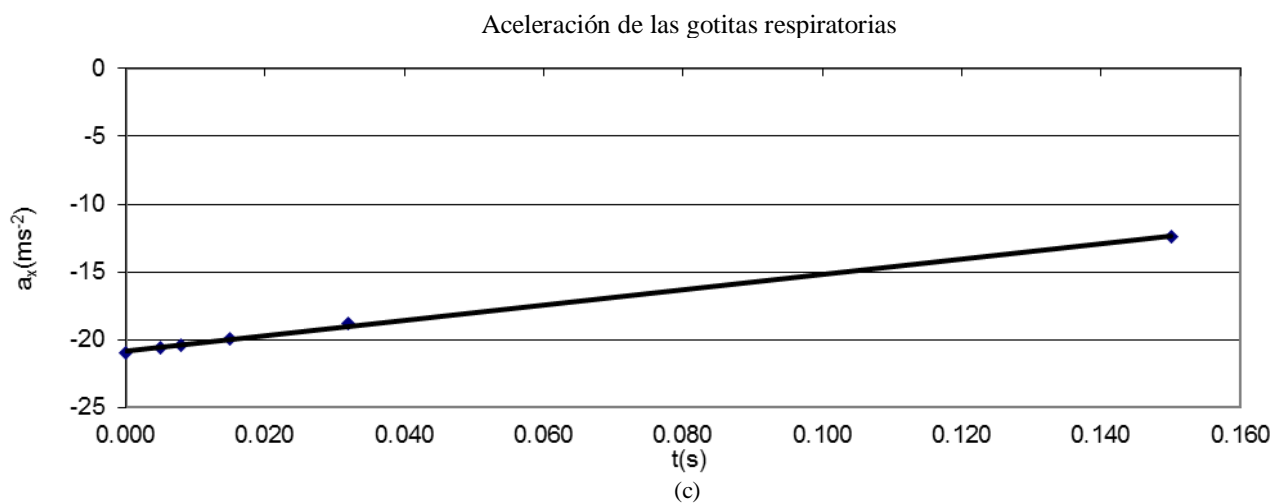

(c)

Figura 4  El desplazamiento, la velocidad y aceleración horizontal en (a), (b) y (c) respectivamente varían exponencialmente en el intervalo de 0 y 0.150 s (Scharfman, BE et al., 2016)

Extrapolando para 2 segundos posteriores a la expulsión de las gotitas respiratorias y aplicando las ecuaciones (9), (10) y (11) se obtiene los siguientes gráficos:

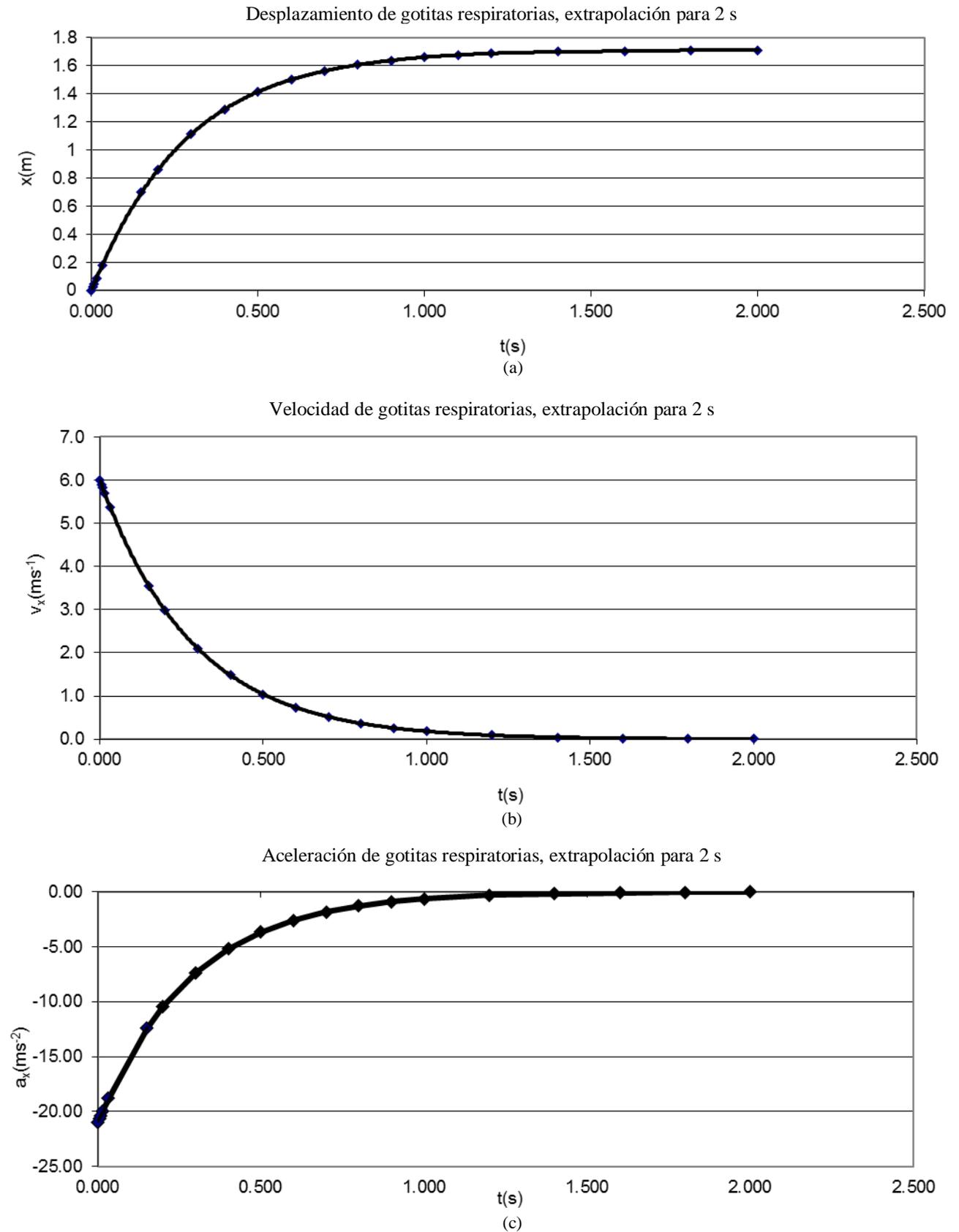

Figura 5 (a) El desplazamiento de la gotita respiratoria alcanza 1.66 m en 1 s, (b) la velocidad y aceleración disminuyen exponencialmente respecto del tiempo

El desplazamiento no cambia sustancialmente en el rango de 1 a 2 s; la velocidad y aceleración disminuyen exponencialmente, en el intervalo de 1 y 2 s tienden a cero.

**3.3 Estimación del desplazamiento en y, velocidad y aceleración de una gotita respiratoria y del nuevo coronavirus.**

En un lugar geográfico a 0 m.s.n.m.; el valor de la aceleración de la gravedad promedio es $g = 9.80 m/s^2$, la viscosidad del aire para una temperatura de 20 ºC (a 1 atm de presión) es $\eta = 1.83 x 10^{-5} N.s/m^2$ (Sears F.W., Zemansky M.W., Young H.D., Freedman R.A, 1999), la densidad del agua $\rho = 10^3 kg/m^3$ y del aire es $\rho` = 1.20 kg/m^3$, por otra parte se tiene la densidad aproximada de un coronavirus COVID-19 es de 1 g/cm$^3$ (Barón, Y.M. et al, 2020), obtenido mediante un análisis indirecto.

Tabla 2 Se muestra la estimación de la velocidad terminal de un coronavirus, gotita resultante del habla, gotitas respiratorias y gota de Flügge en el aire a 0 m.s.n.m.

| Partícula | Diámetro medio de la partícula ($\mu m$) | Densidad ($gk/m^3$) | Velocidad terminal ($ms^{-1}$) |
|---|---|---|---|
| Coronavirus COVID-19 (Barón, Y.M. et al, 2020) | 0.1 | $\approx 10^3$ | 2.97x10$^{-7}$ |
| Gotita resultante del Habla (Morawska L. et al, 2009) | 3.5 – 5 | $10^3$ | 9.10x10$^{-5}$ – 7.43x10$^{-4}$ |
| Gotita respiratoria (World Health Organization, 2020) | 5 | $10^3$ | 7.43x10$^{-4}$ |
| Gotita respiratoria (World Health Organization, 2020) | 10 | $10^3$ | 2.97x10$^{-3}$ |
| Gota de Flügge | 100 | $10^3$ | 1.19 |

La velocidad terminal de un coronavirus tiende a ser cero, es decir podría encontrarse suspendido en el aire en el caso que se encuentre libre de contacto con otras micropartículas; las gotitas resultantes del habla tienen velocidad terminales muy bajas, por ejemplo de 0.7 mms$^{-1}$, no se podría distinguir visualmente su descenso. La gotita respiratoria de 10 μm tiene una velocidad terminal de 2.97 mms$^{-1}$, aproximadamente 3 milímetros por segundo, desciende muy lentamente. Asumiendo la adherencia de un coronavirus COVID-19 en gotitas respiratorias, la velocidad terminal no cambiará en la práctica, por tanto predomina la flotabilidad de las gotitas respiratorias en el aire. Las únicas masas que se precipitan al suelo son las gotas de Flügge que tienen diámetros mayores o iguales a 100 μm.

Se ha estimado la velocidad de caída de microgotas de agua y de un coronavirus COVID-19 a una altitud de 3000 m.s.n.m. a una presión atmosférica de 70108 Pa y una temperatura de $20º C$, la densidad atmosférica a esta altura es $\rho` = 0.85 kg/m^3$, la viscosidad del aire es $\eta = 1.81 x 10^{-5} N.s/m^2$ (1976 Standard Atmosphere Calculator - Digital Dutch)

Tabla 3 Las velocidades terminales dependen del diámetro o tamaño de las microgotas de agua, de igual forma se tiene la velocidad terminal de un virus que correspondería al rango de aerosol.

| Partícula | Diámetro medio de la partícula ($\mu m$) | Densidad ($gk/m^3$) | Velocidad terminal ($ms^{-1}$) |
|---|---|---|---|
| Coronovirus COVID-19 (Barón, Y.M. et al, 2020) | 0.1 | $\approx 10^3$ | 3 x 10$^{-7}$ |
| Gotita resultante del habla ((Morawska L. et al, 2009) | 3.5 - 5 | $10^3$ | 9.20x10$^{-5}$ – 7.51x10$^{-4}$ |
| Gotita respiratoria (World Health Organization, 2020) | 5 | $10^3$ | 7.51x10$^{-4}$ |
| Gotita respiratoria (World Health Organization, 2020) | 10 | $10^3$ | 3.01x10$^{-3}$ |
| Gota de Flügge | 100 | $10^3$ | 1.20 |

Las velocidades terminales obtenidas a una altitud de 3000 m.s.n.m. son aproximadamente iguales a las velocidades encontradas a nivel del mar, estos resultados se deben a la viscosidad del aire en ambos lugares no difieren sustancialmente.

**3.4 Trayectoria de una gotita respiratoria en el plano xy**

Con los resultados obtenidos anteriormente referente a las variables cinemáticas en los ejes **x**, **y** se ha procedido a completar una nueva tabla en función del tiempo: x(t), y(t); de esta forma se ha podido obtener los gráficos de *y(x)* para 2 y 5 s. La trayectoria de una gotita respiratoria en un intervalo de 2 s, es prácticamente una línea horizontal como se observa en la figura 6, el alcance es de 1.71 m en 2 s, en este instante su velocidad y aceleración son cero aproximadamente, la gotita respiratoria se encontraría inmóvil y flotando, la viscosidad impediría que se siga moviendo aún con velocidad constante, es decir, la partícula se encontraría en reposo para las condiciones supuestas. En el caso que se considere un tiempo de 5 segundos, la gotita respiratoria se mantendría flotando y sin posibilidad de poder avanzar en la dirección **x**, esto coincide con nuestra percepción cotidiana, es decir, el aire exhalado no se desplaza más de 1.71 m de manera continua.

En el siguiente gráfico se tiene la trayectoria de la gotita respiratoria de 5 μm el cual describe una curva polinómica en una longitud vertical de 1.5 mm y en una distancia horizontal de 1.71 m para un tiempo de 2. Si se observa a unos metros de distancia se podrá observar la trayectoria como una línea en los rangos descritos anteriormente.

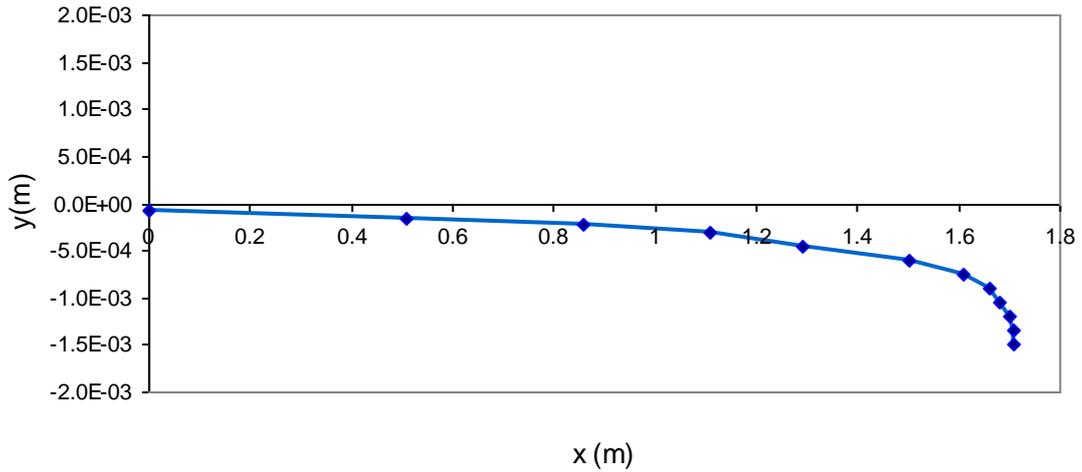

(a)

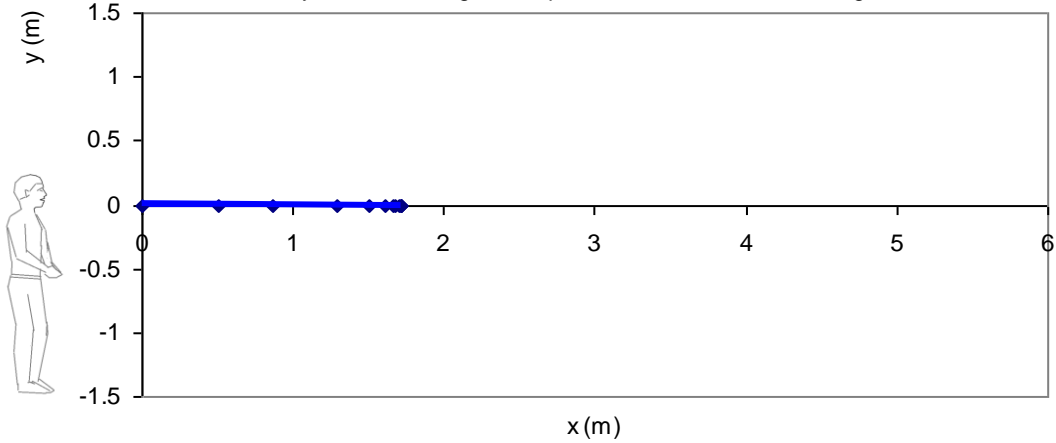

(b)

Figura 6 (a) Trayectoria de una gotita respiratoria en el plano **xy**, la escala en el eje y es de solo milímetros; (b) la trayectoria es una línea si se observa a escala normal.

Empleando las ecuaciones del movimiento para el desplazamiento en el plano **xy**, se tiene el gráfico que permite ver las trayectorias de una gotita respiratoria de 5 μm (color azul) y de una gota de Flügge de 100 μm (color rojo) durante 5 segundos, se puede ver que la trayectoria de una gotita respiratoria es un segmento de línea y queda flotando, mientras la gota de Flügge describe una curva polinómica, la gota impacta sobre la superficie de la Tierra en 5 segundos.

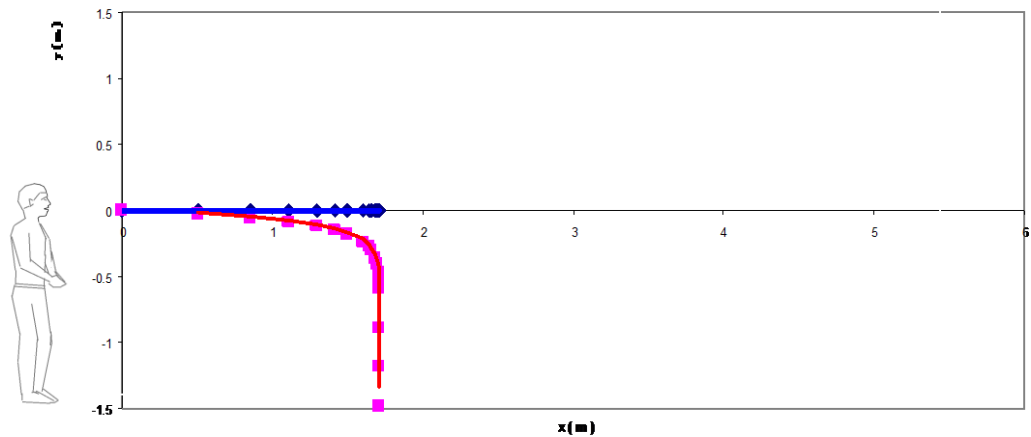

Figura 7 La trayectoria en el plano **xy** de una gotita respiratoria (curva de color azul) y de una gota de Flügge (color rojo), el factor tamaño determina la trayectoria en cada caso.

Si consideramos flujos de aire a partir de 2 s, estos flujos contribuirían a la propagación de las gotitas respiratorias y aerosoles a una distancia aún mayor de 1.71 m y permitiría que alcancen diferentes alturas.

**CONCLUSION**

Se ha determinado que las gotitas respiratorias de 5 – 10 μm flotan en el aire en las cercanías de una persona con síntomas respiratorios, la tos o estornudo de un ser humano conllevaría a que estas gotitas alcancen 1.65 m de distancia en 1 s frenándose rápidamente hasta llegar a 1.71 m en 2 segundos; para 5 segundo, la gotitas respiratorias de 5 -10 μm se encontrarían flotando debido a su velocidad terminal que se encuentra entre $7.43 \times 10^{-4}$ ms$^{-1}$ a $2.97 \times 10^{-3}$ ms$^{-1}$, este rango de velocidades son muy pequeñas para ser detectadas por el ser humano, luego por acción de las condiciones termodinámicas del aire estas microgotas se evaporan alrededor de 7 segundos. La velocidad terminal de gotitas que resulta del habla de 3.5 μm de diámetro (aerosol), es aún mucho menor, de $9.10 \times 10^{-5}$ ms$^{-1}$, estas gotitas también pueden transportar coronavirus COVID-19. Las únicas gotas de agua que caen alrededor de 1.71 m de distancia del paciente con síntomas respiratorios debido al COVID-19, son las gotas de Flügge que logran alcanzar el suelo u otro objeto del medio circundante, por tanto, un coronavirus COVID-19 *se puede desplazar en el aire* adherido a gotitas respiratorias en el entorno de la persona con síntomas respiratorios. La altitud geográfica influye mínimamente en el movimiento de las gotitas respiratorias.